\documentclass[twocolumn,prl]{revtex4-1}
\usepackage{xcolor}
\usepackage{pdfcolmk}
\usepackage{amsmath}
\usepackage{amssymb}
\usepackage{graphicx}
\PassOptionsToPackage{normalem}{ulem}
\usepackage{ulem}
\usepackage[unicode=true,pdfusetitle,
 bookmarks=true,bookmarksnumbered=false,bookmarksopen=false,
 breaklinks=false,pdfborder={0 0 0},pdfborderstyle={},backref=false,colorlinks=true]
 {hyperref}
\hypersetup{
 colorlinks,linkcolor=red,citecolor=blue}

\usepackage{soul}

\makeatletter

\providecolor{lyxadded}{rgb}{0,0,1}
\providecolor{lyxdeleted}{rgb}{1,0,0}

\DeclareRobustCommand{\lyxsout}[1]{\ifx\\#1\else\sout{#1}\fi}

\usepackage{amsfonts}

\makeatother

\begin{document}
\title{Atom-referenced on-chip soliton microcomb}
\author{Rui Niu$^{1,2}$}
\thanks{These authors contributed equally to this work.}
\author{Shuai Wan$^{1,2}$}
\thanks{These authors contributed equally to this work.}
\author{Tian-Peng Hua$^{2}$}
\thanks{These authors contributed equally to this work.}
\author{Wei-Qiang Wang$^{3,4}$}
\thanks{These authors contributed equally to this work.}
\author{Zheng-Yu Wang$^{1,2}$}
\thanks{These authors contributed equally to this work.}
\author{Jin Li$^{1,2}$}
\author{Zhu-Bo Wang$^{1,2}$}
\author{Ming Li$^{1,2}$}
\author{Zhen Shen$^{1,2}$}
\author{Y. R. Sun$^{2,5}$}
\author{Shui-Ming Hu$^{2,5}$}
\author{B. E. Little$^{3,4}$}
\author{S. T. Chu$^{6}$}
\author{Wei Zhao$^{3,4}$}
\author{Guang-Can Guo$^{1,2}$}
\author{Chang-Ling Zou$^{1,2}$}
\email{clzou321@ustc.edu.cn}
\author{Yun-Feng Xiao$^{7}$}
\email{yfxiao@pku.edu.cn}
\author{Wen-Fu Zhang$^{3,4}$}
\email{wfuzhang@opt.ac.cn}
\author{Chun-Hua Dong$^{1,2}$}
\email{chunhua@ustc.edu.cn}
\affiliation{$^{1}$CAS Key Laboratory of Quantum Information, University of Science
and Technology of China, Hefei, Anhui 230026, P. R. China}
\affiliation{$^{2}$CAS Center For Excellence in Quantum Information and Quantum
Physics, University of Science and Technology of China, Hefei, Anhui
230026, People's Republic of China}
\affiliation{$^{3}$State Key Laboratory of Transient Optics and Photonics, Xi\textquoteright an
Institute of Optics and Precision Mechanics, Chinese Academy of Sciences,
Xi\textquoteright an 710119, China}
\affiliation{$^{4}$University of Chinese Academy of Sciences, Beijing 100049,
China}
\affiliation{$^{5}$Department of Chemical Physics, University of Science and Technology of China, Hefei
230026, China}

\affiliation{$^{6}$Department of Physics and Materials Science, City University
of Hong Kong, Kowloon Tong, Hong Kong, China}
\affiliation{$^{7}$State Key Laboratory for Mesoscopic Physics and Frontiers Science
Center for Nano-optoelectronics, School of Physics, Peking University,
Beijing, China}

\begin{abstract}
For the applications of the frequency comb in microresonators, it
is essential to obtain a fully frequency-stabilized microcomb laser source. Here,
we demonstrate an atom-referenced stabilized soliton microcomb generation
system based on the integrated microring resonator. The pump light
around $1560.48\,\mathrm{nm}$ locked to an ultra-low-expansion (ULE) cavity, is frequency-doubled and referenced to the atomic transition of $^{87}\mathrm{Rb}$. The repetition rate of the soliton microcomb is
injection-locked to an atomic-clock-stabilized radio frequency (RF) source, leading
to mHz stabilization at $1$ seconds. As a result, all comb
lines have been frequency-stabilized based on the atomic reference and could be determined
with very high precision reaching $\sim18\,\mathrm{Hz}$ at 1 second,
corresponding to the frequency stability of  $9.5\times10^{-14}$. Our approach provides an integrated and fully stabilized microcomb experiment scheme with no requirement of $f-2f$ technique, which could be easily implemented
and generalized to various photonic platforms, thus paving the way
towards the portable and ultraprecise optical sources for high precision spectroscopy.
\end{abstract}
\maketitle

\section{introduction}

Recently, dissipative Kerr solitons (DKSs) in optical microresonators
have been attracting surging interests \cite{1kippenberg2018dissipative,2herr2014temporal,3wang2020advances,liu2022emerging}
thanks to their self-organizing mechanism that results from the double-balance
between nonlinearity and anomalous dispersion, as well as between
parametric gain and cavity loss. DKSs offer the frequency combs with
high coherence, broad bandwidth and microwave-repetition rate, and
have been applied successfully to optical ranging, dual-comb spectroscopy,
optical clock, quantum key distribution, wavemeter, coherent optical communication
as well as optical frequency synthesis \cite{4ranging,5ranging,6dualcomb,7dualcomb,8dualcomb,9dualcomb,10clock,11quantum,12chaos,13communication,14communication,15microwave,16microwave,17microwave,PhysRevLett.126.063901,tan2021multispecies,obrzud2019microphotonic,35niu,niuNC}.
Among these applications of the soliton microcomb, it is essential
to obtain a fully frequency-stabilized microcomb laser source.

\begin{figure*}[t]
\centerline{\includegraphics[clip,width=1.95\columnwidth]{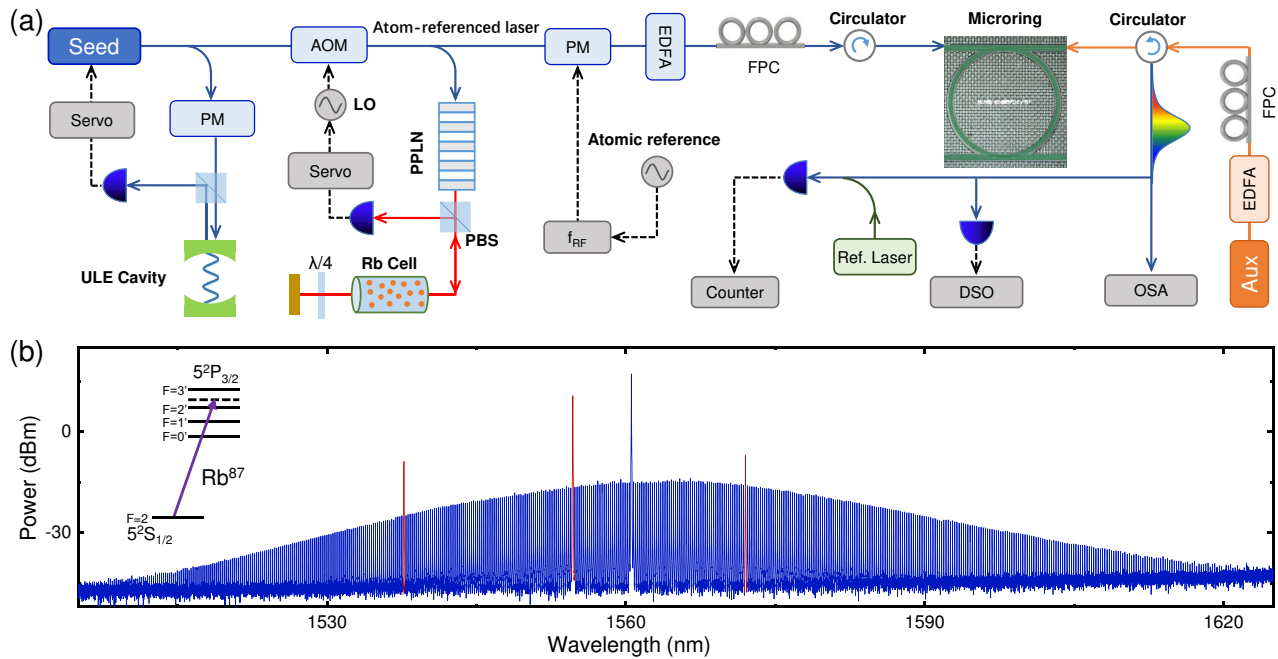}} \caption{\label{fig:Fig1} (a) Schematic of the experimental setup. The frequency of the seed
laser is stabilized by an ULE cavity (finesse of 150000, FSR of 1.5GHz). The atom-referenced laser is generated by shifting the seed laser with an acoustic optical modulator (AOM) and stabilized to the $^{87}\mathrm{Rb}$ D2 transition via periodically
poled lithium niobate (PPLN) frequency
doubler. The repetition rate of the soliton is locked to an atomic-clock-referenced
radio frequency (RF) source via the phase modulator (PM). The pump and
auxiliary lasers are coupled to on-chip microring resonator for the generation of the DKS. Abbreviations: LO, PBS, EDFA, FPC, DSO and OSA are local oscillator, polarization beam splitter, erbium-doped fiber amplifier,
fiber polarization controller, digital oscilloscope and optical spectrum
analyzer, respectively. (b) The optical spectrum of the single soliton
state, and the red lines show the auxiliary laser and the corresponding
optical parametric oscillator lines. Inset: the D2 transition lines,
where the cross-transition of $F'=2$ and $F'=3$ is used to lock the
pump laser as marked by purple arrow.}
\vspace{-6pt}
\end{figure*}

For the soliton microcomb, the mode spectrum $f_{n(\mu)}=f_{\mathrm{ceo}}+n\times f_{\mathrm{rep}}$
$=f_{\mathrm{pump}}+\mu\times f_{\mathrm{rep}}$ ($n$ is the mode number and $\mu$ is the relative mode number with respect to the pump resonance mode) should be stabilized through two degrees
of freedom: the repetition rate $f_{\mathrm{rep}}$ and the carrier envelope
offset frequency $f_{\mathrm{ceo}}$ or pump frequency $f_{\mathrm{pump}}$. However,
these parameters are coupled together with a complex manner by the
microcomb dynamics, and also influenced by several other parameters
in the practical system \cite{18couple}. Consequently, the full stabilization of the microcomb could be realized by controlling the pump laser power, pump laser frequency, and pump resonance detuning,
which requires the reference fiber comb system and several locking
loops \cite{19stable,20locking}. However, the stability of the repetition
rate is limited to the quality of the reference laser, the locking
loop and eventually restricted by the thermal fluctuation in the microcavity
\cite{21noise}. Recently, the concept of injection locking for $f_{\mathrm{rep}}$
has been proposed in the soliton microcomb system \cite{22injection,23injection},
which is realized by modulating the pump laser through a phase modulator
with stabilized radio frequency (RF). This method also provides an approach
to transfer the stabilized RF of the rubidium atomic
clock or the global positioning system (GPS) to the optical frequency
of the soliton microcombs. Furthermore,  with the help of an additional excitation laser around $780\,\mathrm{nm}$, the soliton microcomb can be fully stabilized by locking the comb line to the atomic transition around $1529\,\mathrm{nm}$ using the double-resonance optical pumping technique \cite{24atom}. Such approach often requires multiple reference lasers that produce a large footprint. An integrated microcomb directly referenced to a compact atomic cell \cite{atomcell} may serve the vision to construct a fully integrated optical frequency reference.

In this letter, we demonstrate a fully frequency-stabilized soliton microcomb
with the atomic frequency reference and the injection-locked repetition rate. First, the frequency of the pump laser stabilized by an ultra-low-expansion (ULE) cavity is locked to the absorption line of $^{87}\mathrm{Rb}$ from $5^{2}S_{1/2}$ (F=2) to $5^{2}P_{3/2}$ (F'=(2,3)) with the help of the periodically
poled lithium niobat (PPLN) waveguide. Furthermore, we generate the on-chip soliton microcomb with this stabilized pump laser and demonstrate the injection locking of the repetition
rate with an atomic-clock-stabilized RF oscillator. The linewidth of repetition rate is suppressed from 1.3 kHz to $6.2\,\mathrm{mHz}$ and the frequency stability is also improved to $3.8\times10^{-14}$ at $1\,\mathrm{s}$ of measurement time. The suppress of $f_{rep}$ noise is effective in a large range, which increases with the strength of modulation signal. While the frequency stability of the repetition rate is much better than the pump laser ($\sim18\,\mathrm{Hz}$), the stability of the comb lines over a broadband optical frequency range is directly limited by the pump laser, with the frequency
stability of $9.5\times10^{-14}$ at $1\,\mathrm{s}$ of measurement time. Meanwhile, after the injection locking, the phase noise of the repetition rate also becomes lower than the pump laser, which leads to the entire set of comb lines have the consistent phase noise level. In our system, both the pump laser and the repetition rate are locked to the atomic transition, which allows the frequency of the entire set of comb lines to be stabilized to the absolute frequency of the atomic transition without additional reference lasers. Our approach provides an integrated fully stabilized microcomb experiment scheme with no require of $f-2f$ technique.

\section{Soliton generation}

The experimental setup is shown in Fig.$\,$\ref{fig:Fig1}(a). The microring resonator (MRR) for the soliton microcomb generation is fabricated on
a CMOS-compatible, high-index doped silica glass platform with the loaded
quality (Q) factor of $2.2\times10^{6}$ \cite{31cavity,32cavity}, corresponding to the dissipation rate of $k/2\pi=87\,\mathrm{MHz}$. The integrated dispersion, $D_{\mathrm{int}}\approx\frac{D_{2}}{2}\mu^{2}$, is measured by the fiber Mach-Zehnder interferometer, where $D_{2}/2\pi=27.22\,\mathrm{kHz}$. Firstly, the seed laser is referenced to an ULE cavity (finesse of 150000, FSR of 1.5GHz) through Pound-Drever-Hall (PDH) locking. The short term stability of seed laser is improved while the long term stability is restricted by the cavity drifting (see APPENDIX). For long term stability, the frequency of seed laser is modulated by an acoustic optical modulator (AOM) and locked to the absorption line of $^{87}\mathrm{Rb}$ from $5^{2}S_{1/2}$ (F=2) to $5^{2}P_{3/2}$ (F'=(2,3)) with the help of the PPLN waveguide ~\cite{25xie,26pdh,27pdh}. While the frequency of pump laser is referenced to the ULE cavity and the atomic transition, the short term and long term stability of pump laser are significantly improved, respectively. More details of the stabilized pump laser see APPENDIX.

To get the soliton microcomb,  another auxiliary laser is introduced to balance
the thermal effect and coupled into the MRR in the opposite direction
through the circulator ~\cite{niuNC,29thermalcontrol,30thermalcontrol,zhou2019soliton,34niu,PSC4}. The polarizations of the lasers are controlled
by fiber polarization controllers (FPCs). The lasers are coupled into the MRR with a standard eight-channel fiber array with
a coupling loss of $4.0\,\mathrm{dB}$ per facet, and packaged into
a 14-pin butterfly package with a thermo electric cooler (TEC) chip.
By fine tuning the auxiliary laser when one resonance of the MRR is
coarse tuned close to $f_{\mathrm{pump}}$ by the TEC, the single soliton is realized with the atomic-stabilized laser, as shown in Fig.$\,$\ref{fig:Fig1}(b).
The measured repetition rate
is $26.048\,\mathrm{GHz}$, which agrees well with the size of the MRR.

\section{Experimental RESULTS}

To stabilize the repetition rate, the injection locking of the repetition frequency is realized by modulating the stabilized pump laser with a phase modulator (PM).
Contrary to the free-running soliton, the PM sidebands could trap the soliton by forming intracavity potential gradient in the time domain or discipline the comb lines through the four-wave mixing interaction in the spectrum domain.
When a single soliton is circulating in the microcavity, the intracavity field can be approximately described as
\begin{eqnarray}
\ensuremath{\psi(\theta,t)} & = & \psi_{b}+\psi_{s}
\end{eqnarray}
where $\ensuremath{\psi_{b}}$ and $\psi_{s}$ are the continuous wave background and the characteristic of optical soliton, respectively.
$\theta$ is the rotating angular coordinate.
Through the PM at a modulation frequency $f_{m}$ close to the $f_{rep}$, the total driving term $\tilde{F}$, including the residual pump and generated sideband, are injected into the soliton microcomb.
Therefore, a driving force applied to the soliton via the PM \cite{32cavity} can be rewritten as
\begin{eqnarray}
F & = & \frac{D_{2}}{\kappa\pi}\int_{0}^{2\pi}d\theta\left[\psi^{*}\left(-i\frac{\partial}{\partial\theta}\right)\tilde{F}+c.c.\right] \nonumber \\ & = & -F_{0}\sin\left(\eta\theta_{c}-\Delta\widetilde{\omega}t\right)
\end{eqnarray}
where $F_{0}\approx\frac{8\sqrt{2}\epsilon\eta\sqrt{\left(\omega_{0}-\omega_{p}\right)D_{2}^{3}}}{\pi\kappa^{2}}$ and $\theta_{c}(\tau)$ are the amplitude of F and the center position of the pulse.
$\eta=[f_{m}/f_{rep}]$ is 1 in our experiment and $\epsilon$ presents the modulation strength, where $\epsilon/2$ is the normalized amplitude of modulation sideband.
$\omega_{0}-\omega_{p}$ is the detunning between the pump mode and the pump laser.
$\Delta\widetilde{\omega}=2(2\pi f_{m}- D_{1})/k$ is the normalized detunning between the modulation frequency and angular velocity $D_{1}$. 
The soliton can be presented as a particle with the effective kinetic mass of $M_{0}\thickapprox\frac{2\sqrt{2}}{\pi\kappa}\sqrt{D_{2}\left(\omega_{0}-\omega_{p}\right)}$ when it is unperturbed~\cite{32cavity}.
Based on the momentum analysis, the soliton pulses could move together with the driving force when $F\thickapprox2M_{0}\left|\Delta\widetilde{\omega}\right|<F_{0}$, corresponding to the $\left|\Delta\widetilde{\omega}\right| <  \Delta\widetilde{\omega}_{\mathrm{max}} \thickapprox 2\epsilon\frac{D_{2}}{\kappa}$.
It is noted that the locking range $\Omega_{max}=2\kappa\left|\Delta\widetilde{\omega}_{\mathrm{max}}\right|=2\epsilon{D_{2}}$ is proportional to the modulation strength.

\begin{figure}
\includegraphics[clip,width=1\columnwidth]{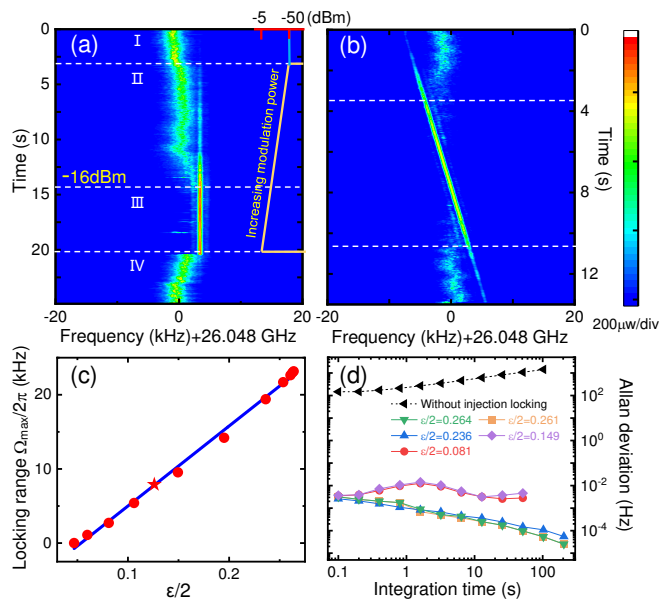}\caption{(a) Typical evolution of the repetition rate when gradually increasing the
modulation strength. (b) Evolution of
the repetition rate around $26.048\,\mathrm{GHz}$ while sweeping the modulation
frequency range from \textbf{$26.0405\,\mathrm{GHz}$} to $26.0537\,\mathrm{GHz}$ with $\epsilon/2=0.125$. $f_{\mathrm{rep}}$ can be synchronized to $f_{\mathrm{m}}$ when
the $\Delta f<2.4\,\mathrm{kHz}$. (c) Locking ranges with
varied modulation strengths. The star shows the result in (b). The blue line shows the linear fitting. (d) The Allan deviation of the repetition
rate for various modulation power. Error bars represent a $68\%$
confidence interval.}
 \label{fig:Fig2} \vspace{-6pt}
\end{figure}

Figure $\,$\ref{fig:Fig2}(a) shows the evolution of the RF
spectrum when the pump laser is modulated
by the PM with various modulation strength. The modulation
frequency is close to the free-spectral range (FSR) of the microresonator and locked to a 10-MHz reference, which is provided by an Hydrogen clock. Without the modulation, the initial repetition rate of the
soliton microcomb is around $26.048\,\mathrm{GHz}$, and the linewidth reaches $\sim1.3\,\mathrm{kHz}$
(stage \mbox{I}). With gradually increasing the modulation strength from
$\epsilon/2=0.001$ to $0.261$ (corresponding to modulation power from $-50\,\mathrm{dBm}$ to $-5\,\mathrm{dBm}$), a sharp peak at $f_{\mathrm{m}}$ appears in the RF spectrum
and the repetition rate gets closer to $f_{\mathrm{m}}$ (stage II), where the modulation power is still relatively weak and the spontaneous generation is still significant, corresponding to the unlocked state. With the increase of PM strength, the intensity
of the initial repetition rate beatnote becomes weaker, indicating
that there exists a competition between these two frequencies. Whereas for large sideband amplitude $\epsilon/2$ and small frequency difference between the sideband and $\omega_{\mathrm{p}}$, the backaction strongly suppresses the spontaneous comb generation process, and drive force-induced generation dominates the parametric process, manifesting the mechanism of injection locking (stage \mbox{III})~\cite{28breath}. Finally,
the repetition rate returns to the initial frequency and the linewidth
becomes as wide as before (stage \mbox{IV}) when the modulation is
turned off. It obviously shows that the phase noise of the system is remarkably suppressed by the significantly narrowed linewidth of the $f_{\mathrm{rep}}$.

Furthermore, the $f_{\mathrm{rep}}$ of DKS can be synchronously tuned by adjusting
the modulation frequency, as shown in the Fig.$\,$\ref{fig:Fig2}(b). We scan the $f_{\mathrm{m}}$ centered around $26.048\,\mathrm{GHz}$ over $14\,\mathrm{kHz}$
with modulation strength of $\epsilon/2=0.125$. During the scanning, when the
absolute frequency difference $\Delta f$ between $f_{\mathrm{m}}$ and initial
repetition rate is relatively large, the modulation signal and the
initial repetition rate can coexist in the RF spectrum. When $\Delta f$
is smaller enough ( $\sim2.4\,\mathrm{kHz}$), $f_{\mathrm{rep}}$ is locked to $f_{\mathrm{m}}$.
We find that after scanning out of the range, the initial repetition
rate drifts to a higher frequency of about $3\,\mathrm{kHz}$, which is attributed
to the modulated pump laser during the measurement. Consequently,
the total locking range can reach about $8\,\mathrm{kHz}$ with $\epsilon/2=0.125$. Figure $\,$\ref{fig:Fig2}(c) shows the
locking range rises monotonically with the modulation strength. The linear fitting (blue curve) slope equals to $107.84\mathrm{kHz}$, which approximately equals to $4D_{2}/2\pi$, consistent with the analytical result.

\begin{figure}
\includegraphics[clip,width=1\columnwidth]{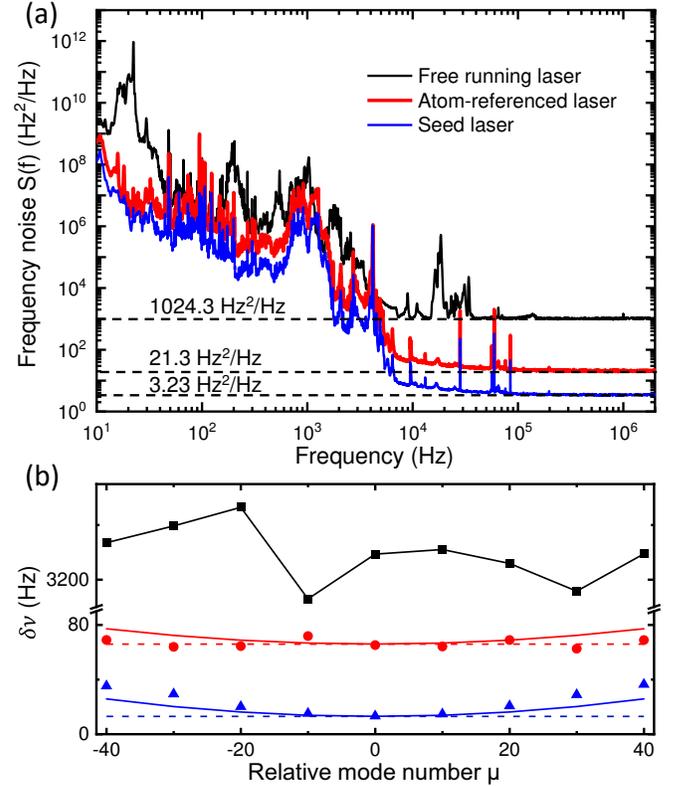}\caption{(a) Single-sided frequency noise spectra of the free running laser, the seed laser, the atom-referenced laser. The corresponding white noises are also indicated. (b) The Lorentz linewidth of comb lines with different $\mu$ (dots). The soliton is generated by the relevant pump lasers in (a) while the repetition rate is injection locked. The red or blue curves shows the theoretical Lorentz linewidth of comb lines with the measured phase noise of repetition rate. The dash-dotted lines represent the white frequency-noise levels of pump laser for guiding the eyes.}
 \label{fig:Fig3} \vspace{-6pt}
\end{figure}

When the repetition rate is injection locked to the RF source, the long term stability of repetition rate should also be improved~\cite{22injection}. Figure $\,$\ref{fig:Fig2}(d) shows the stability of repetition rate with various modulation power. The Allan
deviation is improved
from $146\,\mathrm{Hz}$ to $2\,\mathrm{mHz}$ with $0.1\,s$ of measurement time after the injection locking. When the modulation strength
is lower than $\epsilon/2=0.149$, the long term stability of the system decreases.
With the higher modulation strength ($\epsilon/2=0.236$ $\sim 0.264$),
the long term stability of $f_{\mathrm{rep}}$ is improved to $0.025\,\mathrm{mHz}$ after
$256\,\mathrm{s}$ of measurement time. When the modulation strength is greater
than 0.264, the soliton state is annihilated. This phenomenon
is attributed to the fact that the existence of the soliton state
in the microresonator is broken by the strong intracavity field oscillation
caused by the modulated pump laser, indicating that there is an upper
limitation of the modulation power in our injection locking method, which is different from the previous works using the modulated light for the soliton generation.

The frequency noise and linewidth of individual comb lines
reflecting the coherence performance are essential for
the applications~\cite{camatel2008narrow}. Figure$\,$\ref{fig:Fig3}(a) shows the measured frequency noise spectra of the free running laser, the seed laser and the atom-referenced laser. Here, the frequency noise of comb lines is defined as $S(f)=f^2\times L(f)$, where $L(f)$ is the phase noise of the corresponding comb line and is characterized by the correlated delayed self-heterodyne method~\cite{camatel2008narrow}. It is noted that the frequency noise of laser is distinctly improved when the laser is locked. However, the frequency noise of the atom-referenced laser is degenerated when the seed laser is locked to the atomic transition and the Lorentz linewidth ($\delta\nu$) determined
from the white noise is increased from $10.2\,\mathrm{Hz}$ to $66.9\mathrm{Hz}$. It originates from the seed laser, which is referenced to the broaden atomic transit. Furthermore, the measured Lorentz linewidth of the comb lines with various $\mu$ after the injection locking is shown in Fig. 3(b). Based on the relationship of $L(f_{\mu})=L(f_{\mathrm{pump}})+{\mu}^2L(f_{\mathrm{rep}})$, $L(f_{\mu})$ is dominated by $L(f_{\mathrm{pump}})$ when $L(f_{\mathrm{rep}})$ is much smaller than $L(f_{\mathrm{pump}})$. Thus, the Lorentz linewidth of comb lines is consistent with the pump laser and at the same level. Despite of this, the Lorentz linewidth of comb lines attained sub $\mathrm{kHz}$ level after injection locking. The theoretical Lorentz linewidth of comb lines with the injection locked repetition rate is also shown in curves in Fig.$\,$\ref{fig:Fig3}(b), which indicate that the linewidth increases quadratically with $\mu$ away from the seed laser. When the white frequency-noise of the pump laser is lower, the influence of the $f_{\mathrm{rep}}$ is clearly.

Based on the stabilized pump laser and repetition rate, the frequency of the entire set of comb lines are stabilized to the absolute atomic frequency in our system, which is determined as $f_{\mu}=f_{\mathrm{pump}}+\mu\times f_{\mathrm{rep}}$. In the experiment, the repetition rate is stabilized at mHz level and the pump laser is stabilized at several Hz, thus the frequency stability of comb lines is mainly determined by the pump laser.
To characterize the stability of the fully stabilized comb lines, the wavelength of
a comb line around $1565.368\,\mathrm{nm}$ ($\mu=-23$) is chosen to beat with an reference laser around $1565.368\,\mathrm{nm}$, which is stabilized to another ULE cavity. The reference laser owns the similar parameters with the stabilized seed laser and is locked to a fiber comb system for cancellation of the ULE cavity laser's frequency drift.

\begin{figure}
\centerline{\includegraphics[clip,width=0.95\columnwidth]{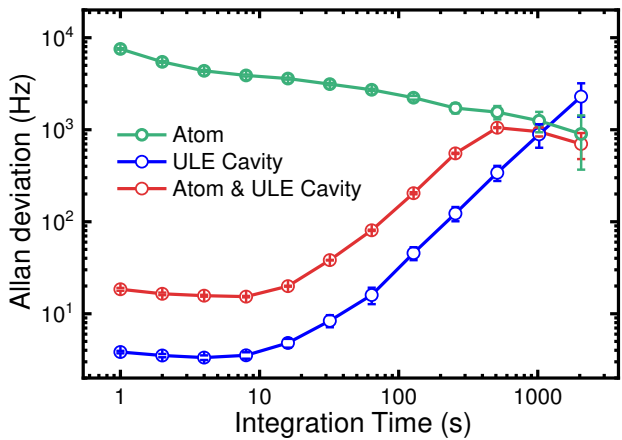}}\caption{\label{fig:Fig4} Comparison of Allan deviation of the soliton generated by the different pump lasers, which are referenced to the atom  transition (green), the ULE cavity (blue), and both atom transition and the ULE cavity (Red), respectively. Error bars represent a $68\%$ confidence interval.
The gate time of the measurement is $100\,\mathrm{ms}$.}
\vspace{-6pt}
\end{figure}
Figure 4 shows the frequency stability of comb lines generated with different pump lasers. It reaches $1.96\times10^{-14}$ ($\sim3.8\,\mathrm{Hz}$) at $1\,\mathrm{s}$ of measurement time when the soliton generated with seed laser referenced to the ULE cavity, as shown the blue curve in Fig.$\,$\ref{fig:Fig4}. However, the result is worse for longer time because of the drifting of the ULE cavity. On the contrary, the stability of the soliton generated by the pump laser, which is only locked to the atomic transition, is better for long term, as shown the green cure in Fig.$\,$\ref{fig:Fig4}. Here, the broaden atomic transition induces the frequency jitter at short term, which is several orders great. Therefore, to improve the stability of the whole system, the seed laser should be referenced to the atomic transition, as shown the red curve in Fig.$\,$\ref{fig:Fig4}. It clearly shows that the red curve has similarly tendency as the blue cure at short term, which means it mainly attribute by the ULE cavity. However, the broaden atomic transition still damaged the stability from $1.96\times10^{-14}$ to $9.5\times10^{-14}$ at $1\,\mathrm{s}$ of measurement time. For long term, the Allan deviation shows an inflection point at $512\,\mathrm{s}$ and drifts at the same rate with the green curve, corresponding to the attribute of the atomic transition. Nevertheless, it can be inferred that the frequency of all comb lines is stabilized at Hz level in our system because of the negligible fluctuation of the repetition rate.

\section{CONCLUSION}

In summary, we demonstrate a fully frequency-stabilized soliton microcomb, by
referencing both the central comb line and repetition rate to atomic
references. By injection locking of the repetition rate when the soliton is generated, the
frequency stability of $f_{\mathrm{rep}}$ reaches $3.8\times10^{-14}$ at 1s of measurement time and the phase noise of the repetition rate is also improved by the RF signal. Based on locking the frequency of pump laser to ULE cavity and atomic transition, the short term and long term stability of the pump laser are significantly improved. Then, the frequency stability of the come line can reach $9.5\times10^{-14}$
at $1\,\mathrm{s}$ of measurement time. Therefore, in our system, the frequencies of all comb lines are stabilized at Hz level. Comparing to previous
 works that required helper lasers ~\cite{24atom,10clock}, or $f-2f$ technique to stabilize $f_{\mathrm{CEO}}$ ~\cite{drake2019terahertz,17microwave}, our work providing a more integrated and scalable approach for full stabilization of the soliton microcomb system, which could be improved by the compact atom cell~\cite{atomcell,hummon2018photonic}, the chip-scale reference cavity~\cite{guo2022chip} and the integrated modulator ~\cite{modulator}. Thus our work holds great potential for integrated precision spectroscopy, optical
clock and optical ranging.

\section*{APPENDIX A}

\begin{figure}[t]
\centerline{\includegraphics[clip,width=0.9\columnwidth]{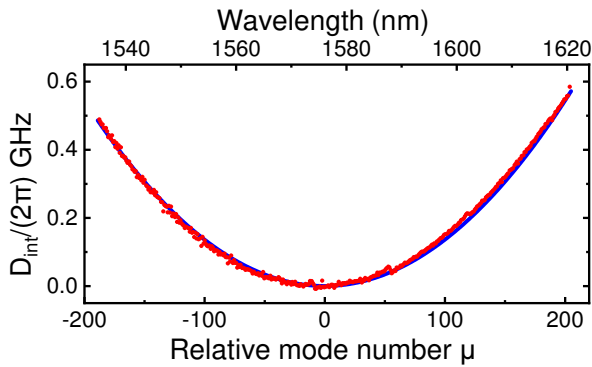} } \caption{\label{fig:FigS6} The measured dispersion of the  integrated cavity.
The red dots and blue lines are the experimental data and the theoretical fitting, respectively.
}
\vspace{-6pt}
\end{figure}

The integrated dispersion is measured by fiber Mach--Zehnder interferometer and shows in Fig.$\,$\ref{fig:FigS6}.
The measured integrated dispersion is $D_{int}=\omega_{\mu}-\omega_{0}-\mu D_{1}\approx\frac{D_{2}}{2}\mu^{2}$, where $D_{2}/2\pi=27.22kHz$, $D_{1}/2\pi=26.048GHz$ and $\omega_{0}/2\pi=192.18THz$.
It is noted that the dispersion of the fiber $-20\,\mathrm{ps}/\left(\mathrm{km}\cdot \mathrm{nm}\right)$ should be considered in the measurement of the integrated dispersion because the dispersion of the $SiON$ is close to the $SiO_{2}$.
And we guess that the difference between the linear result slope $\Omega_{max}/2\pi\epsilon$ and the measured $4D_2/2\pi$ comes from the inaccuracy of the dispersion coefficient, the measurement of the power of the modulation sideband from the optical spectrum and the judgement of the locking range from the electronic spectrum.

\section*{APPENDIX B}
\begin{figure}[h]
\centerline{\includegraphics[clip,width=1\columnwidth]{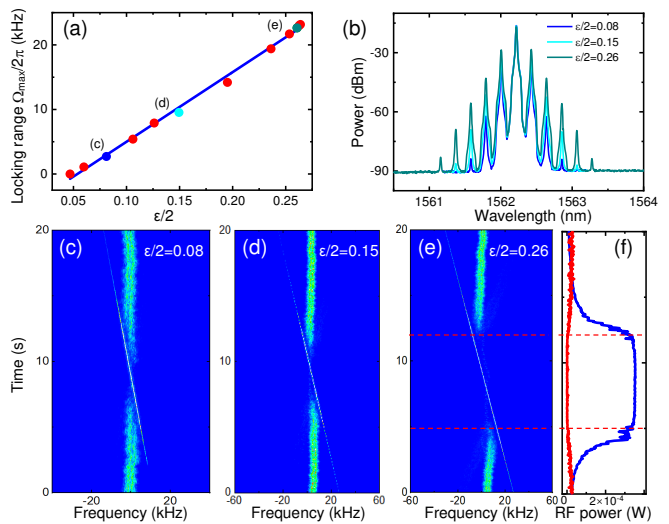} } \caption{\label{fig:FigS3}
(a) Locking ranges with varied modulation strengths (red circles).
And the linear fitting slope $\Omega_{max}/\pi\epsilon$ equals to 107.84kHz, which approximately equals to $4D_{2}/2\pi=108.88$kHz. (b) The typical optical spectrum of modulated pump laser with various modulation powers, corresponding to the modulation strength of $\epsilon/2  = 0.08, 0.15, 0.26$.
(c-e) The typical evolution of the repetition rate around 26.048 GHz while sweeping the modulation frequency with the modulation strength in (b). (f) Evolution of RF power of the modulation frequency (blue curve) and repetition rate (red curve) in (e).}
\vspace{-6pt}
\end{figure}

The repetition rate of the soliton is stabilized by injection
locking. Here, the pump laser is modulated by an phase modulator (PM) with the
modulation frequency $f_{\mathrm{m}}$ near to the repetition rate $f_{\mathrm{rep}}$ of the soliton state.
When gradually increasing the modulation power, several sidebands are
generated near the pump laser, and the sidebands become stronger gradually,
as shown in Fig.$\,$\ref{fig:FigS3}. The repetition rate would be
synchronized to the modulation signal when the modulation power is
strong enough. The effect of injection locking depends on the frequency difference $\Delta f$ of the repetition rate and the modulation frequency. And the locking range increases linearly with the modulation strength, as shown in Fig.$\,$\ref{fig:FigS3} (a). Here, we compare the RF power difference of the modulation signal and the repetition rate to determine whether the system have reached locked state. As shown in Fig.$\,$\ref{fig:FigS3} (f), when the RF power difference above $20\,\mathrm{dB}$, the system is regraded as locked state.

\section*{APPENDIX C}
\begin{figure}[h]
\centerline{\includegraphics[clip,width=1\columnwidth]{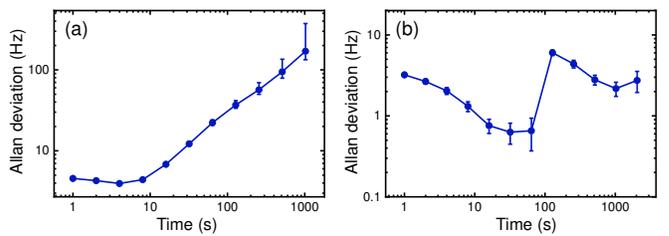} } \caption{\label{fig:FigS1} (a) The Allan deviation of the beat frequency between the locked seed laser and the reference laser. Owning to these two lasers have similar frequency stability, the Allan deviation of the locked seed laser should be divide with $\sqrt{2}$. (b) The Allan deviation of the reference laser after it is referenced to a fiber comb system for cancellation of the ULE cavity's frequency drift.}
\vspace{-6pt}
\end{figure}

The
seed laser (Toptica CTL 1550) is locked to an ultra-low-expansion (ULE) cavity (finesse 150000, linewidth $10\,\mathrm{kHz}$). To characterize the stability of the seed laser, we introduce an reference laser around $1565.368\,\mathrm{nm}$ for measurement. The reference laser is referenced to another ULE cavity with finesse 150000 and linewidth $10\,\mathrm{kHz}$. Therefore, the seed laser is also firstly locked around $1565.368\,\mathrm{nm}$. The measured Allan deviation shows that the locked seed laser owns the stability of $2.8\,\mathrm{Hz}$ at 1s measurement time, as shown in Fig.$\,$\ref{fig:FigS1}(a). Besides, owning to the frequency drift of the cavity, the Allan deviation shows a inflection point at $8\,\mathrm{s}$. We believed that the seed laser around $1560.48\,\mathrm{nm}$ also has the similar stablitiy when it is locked to the ULE cavity.

\begin{figure}[h]
\includegraphics[clip,width=1\columnwidth]{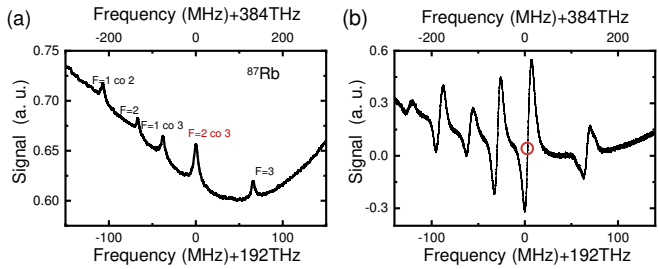} \caption{\label{fig:FigS2} (a) The detected saturated absorption spectrum
of $^{87}\mathrm{Rb}$. (b) The corresponding error signal for PDH locking,
and the locking point is marked as red circle.}
\vspace{-6pt}
\end{figure}

Then, the pump laser is locked to the atomic transition with the help of an acoustic optical modulator (AOM) and a periodically
poled lithium niobate (PPLN) waveguide for longer term stability. The frequency-doubled laser doubly passes a $^{87}\mathrm{Rb}$ gas cell to obtain the saturation
absorption spectrum. As shown in Fig.$\,$\ref{fig:FigS2}(a), the linewidth of atomic transition ($5^{2}S_{1/2}$ (F=2) to  $5^{2}P_{1/2}$ (F'=(2,3))) is around  $3\,\mathrm{MHz}$ due to the halving of the detected laser frequency from $384\,\mathrm{THz}$ to $192\,\mathrm{THz}$. Then, the frequency doubled
laser ($\sim384\,\mathrm{THz}$) is locked to the atomic transition through Pound-Drever-Hall (PDH) locking. The corresponding error signal for PDH locking is shown
in Fig.$\,$\ref{fig:FigS2}(b), and the locking point is marked as
red circle. With these two locking systems, the short-term stability of the pump laser is mainly depended on the ULE cavity and the long-term stability of the pump laser is mainly depended on the atomic transition. To characterize the stability of the pump laser, the reference laser is referenced to a fiber comb system for cancellation of the ULE cavity's frequency drift, and the Allan deviation of the reference laser is shown in Fig.$\,$\ref{fig:FigS1}(b). However, the short-term stability is also disturbed by the broader atomic transition. To minimize the disturbance, the P parameter of Proportion Integration Differentiation (PID) is set relative small.  The short-term stability still degenerates from 3.8Hz to 18Hz at 1 second, which is estimated from the Allan Deviation of the comb line of Fig. 4 in the main text.

\section*{Funding}
This work was supported
by the National Key Research and Development Program of China (Grant
No.2020YFB2205801), Innovation program for Quantum Science and Technology (2021ZD0303203), National Natural Science Foundation of China (Grant
No.12293052, 12293050, 11934012, 12104442 and 92050109), the CAS Project for Young Scientists in Basic Research (YSBR-069), the Fundamental Research Funds for the Central Universities. W.Q.W., and W.F.Z. acknowledge the Strategic
Priority Research Program of the Chinese Academy of Sciences (Grant
No. XDB24030600).

\section*{Acknowledgment}
The authors thank Q. F. Yang, and J. M. Cui. This work was partially carried out at the USTC Center for Micro and Nanoscale Research and Fabrication.

\section*{Disclosures}
The authors declare no conflicts of interest.

\section*{Data Availability}
Data underlying the results presented in this paper are not publicly available at this time but may be obtained from the authors upon reasonable request.


%

\end{document}